\documentclass[twocolumn,english,floatfix,prl]{revtex4}

\usepackage{amssymb}
\usepackage{times}

\makeatletter

\usepackage{graphicx}
\usepackage{amssymb}
\usepackage{bm}

\newcommand{\bra}[1]{\left \langle#1\right|}
\newcommand{\ket}[1]{\left|#1\right \rangle}

\begin{document}

\title{The time-reversal test for stochastic quantum dynamics}

\author{Mark~R.~Dowling}

\author{Peter~D.~Drummond}

\email{drummond@physics.uq.edu.au}

\author{Matthew~J.~Davis}

\author{Piotr~Deuar}

\affiliation{ARC Centre of Excellence for Quantum-Atom Optics, School of Physical
Sciences, University of Queensland, Brisbane, QLD 4072, Australia}



\begin{abstract}
The calculation of quantum dynamics is currently a central issue in
theoretical physics, with diverse applications ranging from ultra-cold
atomic Bose-Einstein condensates (BEC) to condensed matter, biology,
and even astrophysics. Here we demonstrate a conceptually simple method
of determining the regime of validity of stochastic simulations of
unitary quantum dynamics by employing a time-reversal test. We apply
this test to a simulation of the evolution of a quantum anharmonic
oscillator with up to $6.022\times10^{23}$ (Avogadro's number) of particles.
This system is realisable as a Bose-Einstein condensate in an optical
lattice, for which the time-reversal procedure could be implemented
experimentally. 
\end{abstract}
\maketitle
The difficulty of real-time quantum dynamical calculations is caused
by complexity. The computational resources required to directly represent
the Hilbert space of a large quantum system are enormous. This problem
led to Feynman's proposal~\cite{Feynman1982a} to develop quantum
computer hardware for quantum dynamics. In the absence of such devices,
digital computers must be employed for these calculations at present.
Regardless of the hardware or software being utilised, there is a
profound question of how one can check that results are calculated
accurately. This is an especially difficult issue with the time-evolution
of quantum many-body systems, which is one of the central challenges
in theoretical physics. There are few exact solutions, yet results
must be calculated systematically to within a known error in order
to allow theoretical predictions to be tested experimentally.

Stimulated by the success of quantum Monte-Carlo methods in imaginary
time~\cite{Wilson1974a,Creutz1980a,Pollock1984a,Ceperley1995a},
the method used here for real-time quantum dynamics relies on sampling
a probabilistic phase-space representation. Related approaches include
Wigner's classical phase-space representation~\cite{Wig-Wigner},
which was used to develop semi-classical approximations similar to
those for quantum chaos calculations~\cite{Qchaos}, as well as other
classical phase-space \cite{Husimi,Glauber1963b,Gla-P2,Sud-P} representations.
More recent phase-space methods for quantum simulations use a nonclassical
phase-space together with a weight parameter analogous to those used
in path-integrals~\cite{Deuar2002a}. These methods allow quantum
dynamical simulations from first principles without semiclassical
approximations. However, the sampling error can become a limiting
factor.

Fortunately, an important property of time-independent Hamiltonians
is that evolution backward in time is equivalent to evolution forward
in time under a Hamiltonian of the opposite sign. This suggests a
simple yet powerful test that \emph{any} unitary quantum dynamical
simulation must pass. Beginning with a well-defined initial state,
a simulation is evolved for a time period for which we are interested
in the quantum dynamics. The Hamiltonian is then negated and the simulation
evolved again for the same period. For reliable simulation all relevant
initial observables should be recovered.

Phase-space methods utilising quasi-probability distributions lead
one to sample an equivalent set of stochastic differential equations
(SDEs) with random noise terms, and these techniques scale linearly
with the number of modes~\cite{Drummond1980a,Gardiner1999a,Carusotto2001a,Plimak2001a}.
While such methods have been successful for many problems~\cite{Drummond1993a,Drummond1993b},
the sampling errors sometimes grow in time and eventually can become
unmanageable. Similar issues are encountered in simulating classical
chaos, where sensitive dependence on initial conditions leading to
an exponential growth of errors~\cite{Eckmann1995a} can be tested
via time-reversal. However, the use of intrinsically random equations
for time-reversible quantum evolution appears paradoxical. How can
one have time-reversibility in a method which appears to introduce
increasing entropy at each step? It is this question we focus on here,
by showing that this type of time-evolution is in fact completely
reversible due to the storage of information in quantum correlations.

All currently known phase-space methods can be represented in a unified
manner by an expansion of the density operator as \begin{equation}
\hat{\rho}=\int d\vec{\alpha}\, G(\vec{\alpha})\hat{\Lambda}(\vec{\alpha}),\end{equation}
 where $G(\vec{\alpha})$ is a positive distribution function over
the phase-space $\vec{\alpha}$, and $\hat{\Lambda}(\vec{\alpha})$
is an over-complete basis for the Hilbert space~\cite{Corney2003a}.
A variety of techniques can be realised by changing the basis set,
the dynamical equations (which are equivalent under a {}``stochastic
gauge'' symmetry \cite{Drummond2003a}), and the numerical integration
algorithm. We illustrate the time-reversal test for the particular
case of a stochastic gauge simulation~\cite{Deuar2002a,Drummond2003a}.
For this method the phase-space is $\vec{\alpha}=(\underline{\alpha},\underline{\beta},\Omega)$,
which is a $2M+1$ complex dimensional vector containing phase-space
variables $\alpha_{k}$ and $\beta_{k}$ (where $\underline{\alpha}=\{\alpha_{1},\dots,\alpha_{k},\dots,\alpha_{M}\}$,
etc.) for each of $M$ bosonic modes, together with an additional
variable $\Omega$ termed the weight. The operator basis is \begin{equation}
\hat{\Lambda}(\vec{\alpha})=\Omega\bigotimes_{k}\frac{|\alpha_{k}\rangle\langle\beta_{k}^{*}|}{\langle\beta_{k}^{*}|\alpha_{k}\rangle}\,\,,\label{lambda}\end{equation}
 where the coherent state $\ket{\alpha_{k}}$ is an eigenstate of
the boson annihilation operator $\hat{a}_{k}$ for the $k$th mode,
with a mean boson number $\bar{n}_{k}=\langle\hat{a}_{k}^{\dag}\hat{a}_{k}\rangle\,\,=|\alpha_{k}|^{2}$.

For two-body interactions the master equation for time evolution of
the density operator can be shown to be equivalent to a Fokker-Planck
equation for the evolution of the quasi-probability gauge distribution
$G(\vec{\alpha})$ with basis (\ref{lambda}). This in turn is equivalent
to a set of SDEs. The moments of the gauge distribution function are
then equivalent to dynamical quantum averages of products of bosonic
creation and annihilation operators. For a single mode this equivalence
can be expressed (from now on we omit the mode label $k$) as 
\begin{equation}
\langle\left(\hat{a}^{\dagger}\right)^{m}\hat{a}^{n}\rangle_{\mathrm{QM}}=\frac{\langle\Omega\beta^{m}\alpha^{n}+(\Omega\alpha^{m}\beta^{n})^{*}\rangle_{\mathrm{stoch}}}{\langle\Omega+\Omega^{*}\rangle_{\mathrm{stoch}}},
\label{eq:gaugePquantstochav}
\end{equation}
where $\langle\rangle_{\mathrm{QM}}$ indicates a quantum mechanical
average, and $\langle\rangle_{\mathrm{stoch}}$ is a stochastic average.

In this paper we consider the anharmonic oscillator Hamiltonian \begin{equation}
\hat{H}=\frac{\hbar\kappa}{2}\hat{a}^{\dagger2}\hat{a}^{2},\label{eq:H}\end{equation}
 which describes both the Kerr effect in nonlinear optics, and a single
mode Bose-Einstein condensate (BEC). It is perhaps the simplest model
of a many-body quantum system, and, as it is analytically solvable,
it provides an excellent testing ground for simulation methods.

An important quantum feature of this Hamiltonian is that given an
initial coherent state $\hat{\rho}(0)=\ket{\alpha}\bra{\alpha}$,
the dynamics display a series of collapses and revivals. From the
analytic solution it is known that the quantum averages of the quadrature
variables $\hat{X}=(\hat{a}+\hat{a}^{\dagger})/2$ and $\hat{Y}=(\hat{a}-\hat{a}^{\dagger})/2i$
undergo oscillations that eventually damp out to zero. However, after
a certain time the oscillations revive, and the initial state is recovered.
Defining the dimensionless time variable $\tau=\sqrt{\bar{n}}\kappa t/2\pi$,
where $\bar{n}=|\alpha|^{2}$ is the mean particle number, the relevant
time scales are the oscillation period $\tau_{\mathrm{osc}}\sim O(1/\sqrt{\bar{n}})$,
the collapse time $\tau_{\mathrm{coll}}\sim O(1)$, and the revival
time $\tau_{\mathrm{rev}}=\sqrt{\bar{n}}$. For large $\bar{n}$ the
revival time is many times the collapse time, which in turn is much
longer than the natural oscillation period. This revival is a uniquely
quantum feature that does not occur in classical dynamics.

While single mode Hamiltonians are often not a good description of
real systems, the anharmonic oscillator can be a good approximation
for a Bose-Einstein condensate (BEC) in an optical lattice in the
Mott insulator regime~\cite{Greiner2002a}. A sudden increase in
the lattice depth from the superfluid regime can create coherent superpositions
of atoms at each site which can be approximated by a coherent state.
Indeed, such collapses and revivals have been observed with a BEC
in a deep lattice~\cite{Greiner2002b}.

With a suitable choice of stochastic gauge representation~\cite{Deuar2002a}
it was found to be possible to simulate past the collapse time with
small statistical error using a modest number ($\sim10^{4}$) of stochastic
trajectories~\cite{Drummond2003a}. Here we check this calculation
with the time-reversal test to demonstrate that the full quantum nature
of the dynamics is preserved, even when the mean quadrature amplitudes
are near zero.

One finds \cite{Drummond2003a} that the Ito SDEs corresponding to
the anharmonic oscillator Hamiltonian (\ref{eq:H}) are 
\begin{eqnarray}
\dot{\alpha} & = & i\alpha\left[ -\kappa n_{x}+\xi_{g,1}\right] , \\
\dot{\beta} & = & i\beta\left[ \kappa n_{x}+\xi_{g,2}^{*}\right] , \\
\dot{\Omega} & = & -\Omega n_{y}e^{-g}\sqrt{i\kappa}\left[\xi_{1}-i\xi_{2}\right],\end{eqnarray}
 where $n=n_{x}+in_{y}=\alpha\beta$ is a complex variable corresponding
to the particle number. Here $\xi_{g,j}$ are defined as transformations
of the fundamental noise terms $\xi_{j}$ through the introduction
of an arbitrary stochastic \emph{diffusion gauge} $g$, chosen for
efficiency:
\begin{equation}
\xi_{g,j}(t)=\sqrt{i\kappa}\left[\xi_{j}\cosh g+i\xi_{3-j}\sinh g\right].
\end{equation}
For numerical integration with time steps $dt$, the $\xi_{j}$ can
be implemented by independent real Gaussian noises of variance $1/dt$
and mean zero at each time step. The drift gauge used to obtain the
deterministic parts of the equations from the Hamiltonian has been
described previously~\cite{Drummond2003a}. It is convenient to transform
these equations to logarithmic variables, and to use the Stratonovich
calculus for integration~\cite{Gardiner1999a}. To obtain the time-reversed
SDEs we simply replace $\kappa$ with $-\kappa$ in the above equations,
and generate new, uncorrelated noises.

\begin{figure}
\begin{center}\includegraphics[%
  width=7cm]{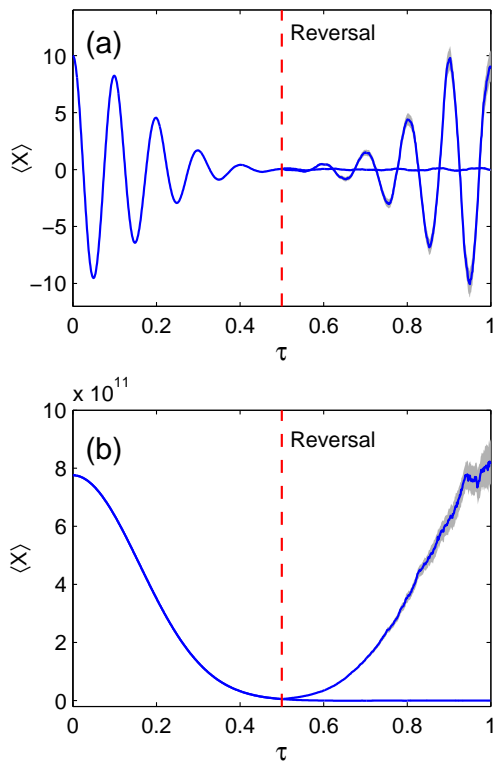}\end{center}

\vspace*{-0.5cm}

\caption{(Color online) Time-reversal test for a stochastic gauge simulation of the $X$-quadrature
of the anharmonic oscillator. (a) An initial coherent state with mean
boson number $\bar{n}=100$ and $10^{5}$ stochastic trajectories.
(b) An initial coherent state with $\bar{n}=6.022\times10^{23}$ and
$10^{7}$ stochastic trajectories, with the calculation performed in the rotating
frame. The solid lines are the simulation
result for $\langle\hat{X}\rangle$, with the statistical error bars
(representing one standard deviation) shown by the grey shading. For
both cases the time-reversal was implemented at $\tau_{R}=0.5$ by
negating the sign of the Hamiltonian and the initial state is recovered
to within statistical error at $\tau=2\tau_{R}$. The forward time
evolution to $2\tau_{R}$ demonstrating the collapse to $\langle\hat{X}\rangle$
= 0 is also shown in both (a) and (b).\vspace*{-0.5cm}}

\label{fig:cgaugerevN100}
\end{figure}

Figure~\ref{fig:cgaugerevN100} illustrates the time-reversal test
for the calculation of the dynamics of the quantum average of the
$X$-quadrature. The initial coherent state has $\bar{n}=\langle\hat{a}^{\dag}\hat{a}\rangle=100$,
and each stochastic trajectory was evolved forward in time until $\tau_{R}=0.5$,
using a constant diffusion gauge of $g=1.6$. The Hamiltonian was
then negated and the system evolved again for the same period. We
observe a reversal in the $X$-quadrature dynamics back to its initial
value to within statistical error, even though each stochastic trajectory
has evolved with uncorrelated noises at every point in the time-evolution.
Of course, there are random features in each trajectory that are not
time-reversed, but these change the distribution in a way that does
not affect observables. This remarkable property is due to the overcompleteness
of the quantum mechanical basis of coherent states, which permits
the same physical state to be represented in more than one way in
terms of coherent states.

While certainly not small, a one-hundred dimensional Hilbert space
is accessible with current computers. We have therefore repeated this
calculation for a much larger mean boson number equal to Avogadro's
number $\bar{n}=6.022\times10^{23}$ --- a truly macroscopic number
of particles. This requires us to make use of more sophisticated gauge
methods, with \begin{equation}
g(\tau,\vec{\alpha})=\frac{1}{6}\log\left\{
\frac{8\pi}{\sqrt{\bar{n}}}|n(\tau)|^{2}[2\tau_{R}-\tau]+\left[1+4n_{y}^{2}(\tau)\right]^{3/2}\right\}.
\end{equation}
 Further details on choices of gauges will be published elsewhere,
also see \cite{Deuar2004}.

Here the total Hilbert space dimension is astronomically large, and
well beyond the capacity of any known or planned digital computer.
The dynamical evolution obtained from the analytic result is a Gaussian
amplitude decay, quite different to the usual exponential decay of
a damped system. While the amplitude decay appears to correspond to
information loss, in fact there is information stored in quantum correlations,
which can be recovered through time-reversal.

In this situation the physical collapse time is very short indeed
--- orders of magnitude less than the revival time --- and there are
many oscillations of the $X$-quadrature. We therefore perform the
calculation in a rotating frame, and the envelope of the oscillations
is plotted in Fig.~\ref{fig:cgaugerevN100}(b). The time reversal
is implemented at $\tau_{R}=0.5$, and again we can see the revival
of the initial state. While this situation is somewhat idealised,
it demonstrates a fully quantum calculation for a macroscopic particle
number.

\begin{figure}
\begin{center}\includegraphics[%
  width=7cm]{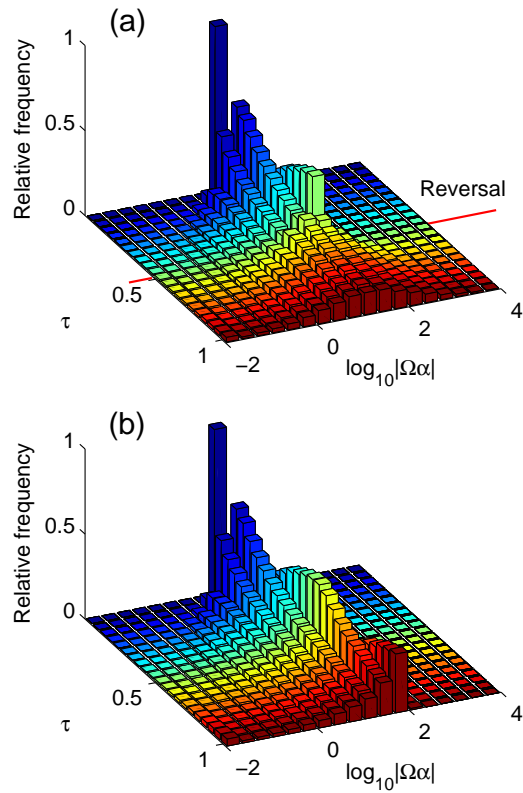}\end{center}

\vspace*{-0.5cm}

\caption{(Color online) A representation of the broadening of the gauge distribution function
for $\bar{n}=100$. The plots are histograms of the quantity $\log_{10}|\alpha\Omega|$
for the stochastic trajectories at a number of points in time. (a)
The time-reversed case with the Hamiltonian negated at $\tau_{R}=0.5$
and further evolved to $\tau=2\tau_{R}$. (b) The forward time evolution
to $\tau=2\tau_{R}$. The initial distribution at $\tau=0$ for both
graphs is a delta function, however it broadens with time to span
several orders of magnitude at $\tau=2\tau_{R}$, despite being another
representation of the initial quantum state in (a). Note that the
\emph{logarithm} of $|\alpha\Omega|$ is binned. \vspace*{-0.5cm}}

\label{fig:diffuse}
\end{figure}

The important result to note is that in both cases the time-reversed
quadrature mean agrees with the initial value to within the sampling
error-bars. The error-bars can be reduced by including more trajectories
in the calculation.

For these calculations the time-reversal test illustrates a powerful
yet counterintuitive feature of stochastic simulations --- they can
be useful for simulating unitary (reversible) quantum dynamics, despite
the irreversible nature of stochastic processes. The examples demonstrate
a stochastic simulation of a quantum revival, a uniquely unitary feature.
As discussed previously, the quadrature variables will display a true
revival to their initial values at $\tau_{\mathrm{rev}}=\sqrt{\bar{n}}$.
By time-reversing the calculation after the initial collapse we induce
the revival early. However, the stochastic trajectories themselves
continue to diffuse under time-reversal --- they do not simply retrace
their forward-time path. Hence, although it seems natural to associate
irreversible stochastic process with irreversible quantum dynamics
(such as those of an open quantum system), clearly this intuition
is unnecessarily restrictive.

The calculation can equivalently be discussed in terms of the distribution
function. During both the forwards and backwards time dynamics the
gauge distribution $G(\vec{\alpha})$ evolves according to a Fokker-Planck
equation with positive-definite diffusion and therefore can only broaden
in phase space. This is clearly illustrated in Fig.~\ref{fig:diffuse}.
The gauge distribution function that is recovered after the reversal
in Fig.~\ref{fig:diffuse}(a) is not the same as the initial condition,
but is still equivalent to the original density operator. This final
distribution is less compact than the original, but nonetheless will
have identical moments corresponding to the normally-ordered operator
averages of the initial state. For this example we have sampled a
gauge distribution $G(\vec{\alpha})$ at $2\tau_{R}$ that is equivalent
to the initially chosen delta function $G(\vec{\alpha})=\delta^{2}(\alpha-\beta^{*})\delta^{2}(\Omega-1)\delta^{2}(\alpha-\sqrt{\bar{n}})$.
This is precisely what must happen at the true revival time $\tau=\sqrt{\bar{n}}$.
Fig.~\ref{fig:diffuse}(b) shows the behaviour of the distribution
function for the forward time evolution for comparison.

Although the calculations are presented as a method of testing quantum
simulations, an early revival could be observed experimentally with
a BEC in an optical lattice using the phenomenon of a Feshbach resonance~\cite{Cornish2000a,Weber2003a}.
This allows the tuning of both the magnitude and sign of the interaction
strength in atomic Bose gases, represented by our parameter $\kappa$,
using an applied magnetic field. As the setup of Greiner \emph{et
al.}~\cite{Greiner2002b} uses only optical potentials for the observation
of revivals, an early revival experiment could be performed simply
with the addition of a precisely-controlled homogeneous magnetic field.

We note that while the calculation presented is for a quantum phase
space method, the time-reversal test is applicable to any quantum
simulation technique. It is not a sufficient test in itself, since
a time-reversible simulation could have other systematic errors. However,
it has the great advantage that time-reversibility is an exact property
of unitary quantum dynamics even when no other exact properties are
known. We believe that the current status of calculating the dynamics
of quantum many-body systems is similar to the situation in the early
days of studying classically chaotic systems on a computer. By definition
such systems display sensitive dependence on initial conditions, and
so it is difficult to estimate the errors in the calculated dynamics~\cite{Eckmann1995a}.
It was partly due to time-reversal tests that such calculations became
convincing.

In summary, we have presented a simple yet powerful test for unitary
quantum dynamics, and demonstrated its use to verify the results of
a stochastic numerical simulation in a macroscopically large Hilbert
space. Such tests are crucial for these demanding calculations. An
increasing variety of quantum dynamical techniques are now becoming
available, and it is important to have reliable tests of their accuracy
--- especially since no analytic solutions exist in many cases of
interest.

We gratefully acknowledge research support from the Australian Research
Council.

\bibliographystyle{apsrev}

\end{document}